\let\ifapj\iffalse
\let\ifarxiv\iftrue
\let\iflocal\iffalse
\ifapj\documentclass[manuscript, linenumbers]{aastex7}\fi
\ifarxiv\documentclass[twocolumn]{article}\fi
\iflocal\documentclass[twocolumn]{article}\fi
\ifapj\usepackage{patch-apj}\NewPageAfterKeywords\fi
\usepackage{etoolbox,ifxetex,ifluatex}
\ifluatex\AtEndPreamble{\hypersetup{pdfencoding=auto}}\fi
\ifboolexpr{bool{arxiv} or bool{local}}{
  \usepackage{font-termes}
}{}
\ifboolexpr{bool{xetex} or bool{luatex}}{
  \newfontfamily\scfont{Noto Serif SC}
  \DeclareTextFontCommand\textscfont\scfont
}{
  \usepackage[utf8]{inputenc}
  \usepackage{CJKutf8}
  \newcommand*\textscfont[1]{\begin{CJK*}{UTF8}{gbsn}#1\end{CJK*}}
}
\usepackage{csquotes,hyperref}
\ifapj
  \usepackage{appendix}
  \let\textacsize\relax
  
  \let\symcal\mathcal
\fi
\makeatletter\ifboolexpr{bool{arxiv} or bool{local}}{
  \usepackage{draft}
  \usepackage[physics, astro]{basic}
  \usepackage[mode=biblatex, colorlinks]{apjbib}
  \ExecuteBibliographyOptions{embedlinks}
  \addbibresource{minidisk.bib}
  \hypersetup{urlcolor=blue}
}{}\makeatother
\usepackage{acro}
\acsetup{single}
\NewAcroCommand\iacp{m}
  {\acroindefinite\acroplural\UseAcroTemplate{first}{#1}}
\NewAcroCommand\Iacp{m}
  {\acroupper\acroindefinite\acroplural\UseAcroTemplate{first}{#1}}
\NewAcroCommand\iaclp{m}
  {\acroindefinite\acroplural\UseAcroTemplate{long}{#1}}
\NewAcroCommand\Iaclp{m}
  {\acroupper\acroindefinite\acroplural\UseAcroTemplate{long}{#1}}
\DeclareAcronym{MBH}{
  short           =MBH,
  long            =massive black hole,
  short-indefinite=an}
\DeclareAcronym{MBHB}{
  short           =MBHB,
  long            =massive black hole binary,
  short-indefinite=an,
  long-plural-form=massive black hole binaries}
\DeclareAcronym{AGN}{
  short           =AGN,
  long            =active galactic nucleus,
  short-indefinite=an,
  long-indefinite =an,
  long-plural-form=active galactic nuclei}
\DeclareAcronym{GW}{
  short           =GW,
  long            =gravitational wave}
\DeclareAcronym{EM}{
  short           =EM,
  long            =electromagnetic,
  short-indefinite=an,
  long-indefinite =an}
\DeclareAcronym{UV}{
  short           =UV,
  long            =ultraviolet,
  long-indefinite =an}
\DeclareAcronym{IR}{
  short           =IR,
  long            =infrared,
  long-indefinite =an}
\DeclareAcronym{MHD}{
  short           =MHD,
  long            =magnetohydrodynamic,
  short-indefinite=an,
  short-plural    =,
  long-plural     =s}
\DeclareAcronym{RMHD}{
  short           =RMHD,
  long            =radiative magnetohydrodynamic,
  short-indefinite=an,
  short-plural    =,
  long-plural     =s}
\DeclareAcronym{RT}{
  short           =RT,
  long            =radiative transfer,
  short-indefinite=an}
\DeclareAcronym{MRI}{
  short           =MRI,
  long            =magnetorotational instability,
  short-indefinite=an}
\DeclareAcronym{ISCO}{
  short           =ISCO,
  long            =innermost stable circular orbit,
  short-indefinite=an,
  long-indefinite =an}
\DeclareAcronym{NASA}{
  short           =NASA,
  long            =National Aeronautics and Space Administration,
  short-indefinite=an}
\DeclareAcronym{NSF}{
  short           =NSF,
  long            =National Science Foundation,
  short-indefinite=an}
\DeclareAcronym{LISA}{
  short           =LISA,
  long            =Laser Interferometer Space Antenna}
\DeclareAcronym{PTA}{
  short           =PTA,
  long            =pulsar-timing array}
\DeclareAcronym{ZTF}{
  short           =ZTF,
  long            =Zwicky Transient Facility}
\DeclareAcronym{LSST}{
  short           =LSST,
  long            =Large Synoptic Survey Telescope,
  short-indefinite=an}
\DeclareAcronym{ULTRASAT}{
  short           =ULTRASAT,
  long            =Ultraviolet Transient Astronomy Satellite,
  long-indefinite =an}
\usepackage{siunitx}
\newcommand*\mH{m_{\element H}}
\newcommand*\eff{_{\su{ef{}f}}}
\usepackage{xcolor}
\expandafter\def\csname editcolor1\endcsname{blue!60!cyan}
\ifapj\let\edit\relax\fi
\newcommand*\edit[2]{\textcolor{\csname editcolor#1\endcsname}{#2}}
\begin{document}

\title{Radiative magnetohydrodynamics simulation of minidisks in equal-mass
massive black hole binaries}

\ifapj
  \author[0000-0001-5949-6109]{Chi-Ho Chan}
  \affiliation{Center for Relativistic Astrophysics and School of Physics,
  Georgia Institute of Technology, Atlanta, GA 30332, USA}
  \email{chchan@gatech.edu}
  \author[0000-0002-7110-9885]{Vishal Tiwari}
  \affiliation{Center for Relativistic Astrophysics and School of Physics,
  Georgia Institute of Technology, Atlanta, GA 30332, USA}
  \email{vtiwari@gatech.edu}
  \author[0000-0002-7835-7814]{Tamara Bogdanovi\'c}
  \affiliation{Center for Relativistic Astrophysics and School of Physics,
  Georgia Institute of Technology, Atlanta, GA 30332, USA}
  \email{tamarab@gatech.edu}
  \author[0000-0002-2624-3399]{Yan-Fei Jiang (\textscfont{姜燕飞})}
  \affiliation{Center for Computational Astrophysics, Flatiron Institute,
  162 Fifth Avenue, New York, NY 10010, USA}
  \email{yjiang@flatironinstitute.org}
  \author[0000-0001-7488-4468]{Shane W. Davis}
  \affiliation{Department of Astronomy, University of Virginia,
  Charlottesville, VA 22904, USA}
  \email{swd8g@virginia.edu}
\fi

\ifboolexpr{bool{arxiv} or bool{local}}{
  \author{Chi-Ho Chan}
  \author{Vishal Tiwari}
  \author{Tamara Bogdanovi\'c}
  \affil{Center for Relativistic Astrophysics and School of Physics,
  Georgia Institute of Technology, Atlanta, GA 30332, USA}
  \author{Yan-Fei Jiang (\textscfont{姜燕飞})}
  \affil{Center for Computational Astrophysics, Flatiron Institute,
  162 Fifth Avenue, New York, NY 10010, USA}
  \author{Shane W. Davis}
  \affil{Department of Astronomy, University of Virginia,
  Charlottesville, VA 22904, USA}
}{}

\date{May 3, 2025}
\ifapj\else\uats{%
  Radiative magnetohydrodynamics (2009);
  Supermassive black holes (1663);
  Gravitational wave sources (677);
  Accretion (14);
  Black hole physics (159);
  Gravitation (661)}
\fi

\shorttitle{RMHD minidisks in equal-mass MBHBs}
\shortauthors{Chan et al.}
\pdftitle{Radiative magnetohydrodynamics simulation of minidisks in equal-mass
massive black hole binaries}
\pdfauthors{Chi-Ho Chan, Vishal Tiwari, Tamara Bogdanovi\'c, Yan-Fei Jiang,
Shane W. Davis}

\begin{abstract}
We are on the cusp of detecting \acp{GW} from individual \acp{MBHB} with the
\acl{LISA} and \aclp{PTA}. These \acp{MBHB} may be surrounded by circumbinary
disks and minidisks, the \acl{EM} emission from which are essential for
localizing the \acp{MBHB} on the sky. Here we present the first \acp{RMHD}
minidisk simulation that directly solves the \acl{RT} equation on discretized
grid rays. The simulation examines one of the minidisks in an equal-mass
\qty{2e7}{\solarmass} \ac{MBHB} separated by 100 gravitational radii. Minidisks
simulated with and without radiative effects resemble each other qualitatively
but differ in several key aspects. The \acp{RMHD} minidisk is denser and
geometrically thinner than the \aclp{MHD} minidisk. Furthermore, the \acp{RMHD}
minidisk, with a nonaxisymmetric photosphere and temperature distribution,
produces an anisotropic illumination pattern. As a result, the observed
radiative flux of two \acp{RMHD} minidisks orbiting each other varies at half
the binary orbital period, a feature independent of relativistic boosting and
lensing effects. Such periodic light curves, if identified in upcoming optical
transient surveys, could reveal the existence of \acp{MBHB} on the way to
merger, particularly if they are in a constant phase relation with detected
\acp{GW}.
\end{abstract}
\acresetall

\section{Introduction}

Detecting \acp{GW} from \acp{MBHB} is fast becoming a reality. \Acp{PTA} such
as \textacsize{NANO}Grav have discovered a low-frequency \ac{GW} background
that is consistent with it being produced by a population of \acp{MBHB}
\citep[e.g.,][]{2023ApJ...951L...8A, 2023ApJ...952L..37A}. \Acp{PTA} have also
performed targeted searches for individual \acp{MBHB}
\citep[e.g..][]{2023ApJ...951L..50A, 2024ApJ...963..144A}. These tantalizing
observational developments fuel a renewed sense of urgency to better understand
these \acp{MBHB} theoretically. The anticipated launch of \ac{LISA} in about a
decade \citep{2017arXiv170200786A, 2023LRR....26....2A, 2024arXiv240207571C}
provides similar motivation.

For any \acp{MBHB} to be detected at all, close \ac{MBH} pairs must result from
at least a fraction of galaxy mergers. Additionally, for some of these
\acp{MBHB} to be caught in the act of coalescence, their merger times must be
shorter than the age of the Universe. Observations of dual and multiple
\acp{AGN} with $\mathrelp\sim\unit{\kilo\parsec}$ orbital separations confirm
that galaxy mergers are natural sites for forming wide \ac{MBH} pairs and even
multiples \citetext{see \citealp{2019NewAR..8601525D} for a review}. Recent
\ac{PTA} evidence for \iac{GW} background attributable to \acp{MBHB} further
suggests that subparsec \acp{MBHB} do form and merge within a Hubble time.

Observationally connecting galaxy mergers with \ac{MBH} mergers remains
challenging because the two kinds of mergers happen \qtyrange{\sim
e8}{e9}{\year} apart in time, and because \ac{GW}-driven inspiral and
coalescence proceed on timescales much shorter than other \ac{MBHB}
evolutionary stages. \Acp{MBHB} headed for merger are difficult to identify in
\ac{EM} observations alone without foreknowledge of how they appear or when and
where they occur \citep[e.g.,][]{2022LRR....25....3B, 2023arXiv231016896D}.
\Acp{GW} unmistakably herald \acp{MBHB} on course to merger, but a source can
only be poorly localized during early inspiral. \Iac{EM} counterpart spatially
coincident with \iac{GW} detection may offer the best chance of singling out
the host galaxy of \iac{MBHB}.

Because individual, subparsec \acp{MBHB} have not been definitively observed,
theoretical studies have been crucial for our understanding of the accretion
flow in \acp{MBHB}, including its structure and emission properties. Such
systems have been extensively studied using hydrodynamics and \acp{MHD}
simulations in Newtonian gravity \citep[e.g.,][]{1994ApJ...421..651A,
2005ApJ...630..152E, 2008ApJ...672...83M, 2009MNRAS.393.1423C,
2012ApJ...749..118S, 2013MNRAS.436.2997D, 2016ApJ...827...43M,
2020ApJ...900...43T, 2021ApJ...909L..13Z, 2023MNRAS.526.5441K,
2023MNRAS.518.5059S, 2023MNRAS.522.2707S, 2024MNRAS.528.2358C,
2024ApJ...970..156D, 2024arXiv241023264M}, approximate general relativity
\citep{2012ApJ...755...51N, 2021ApJ...922..175N, 2015PhRvD..91b4034Z,
2017ApJ...835..199R, 2017ApJ...838...42B, 2018ApJ...853L..17B,
2019ApJ...879...76B, 2021ApJ...913...16L, 2022ApJ...928..187C,
2023MNRAS.520.1285M, 2024ApJ...974..242A, 2025arXiv250206389E}, and full
general relativity \citep{2010ApJ...715.1117B, 2012ApJ...744...45B,
2011PhRvD..84b4024F, 2012PhRvL.109v1102F, 2014PhRvD..89f4060G,
2014PhRvD..90j4030G, 2018PhRvD..97d4036K, 2021ApJ...910L..26P,
2023MNRAS.520..392B, 2023PhRvD.108l4043R, 2024PhRvD.109j3024F}. These
simulations established that, if the gas cools efficiently enough to collect
into a rotationally supported, geometrically thin circumbinary accretion disk,
binary torques can evacuate gas from the central region around the \ac{MBHB}
\citetext{see \citealp{1979MNRAS.188..191L} and the references above}. Each
\ac{MBH} can nevertheless retain a minidisk gravitationally bound to it by
capturing gas from the inner rim of the circumbinary disk. Critically,
accretion across the evacuated cavity continues unhindered despite strong
binary torques, at rates comparable to disks around single \acp{MBH}. This has
two important implications: Angular momentum is efficiently removed from
material crossing the cavity, and binary \acp{MBH} can be powered by accretion
to shine as \acp{AGN} just like single \acp{MBH}.

Earlier theoretical work adopted simple prescriptions for the thermodynamics of
circumbinary disks and minidisks. This was necessary because disk cooling,
governed by radiative processes on the atomic level and the complex interplay
between gas and radiation, is computationally expensive to model from first
principles. However, the impact of radiation on the accretion flow close to the
\ac{MBHB} can be non-trivial. Radiation alters gas density, temperature, and
opacity. Opacity in turn dictates how quickly and at what frequencies radiation
is emitted, absorbed, and scattered. The tight gas--radiation coupling demands
that both are evolved simultaneously with \acp{RMHD} simulations, particularly
if reliable observational predictions of circumbinary disks and minidisks are
to be made based on these simulations.

\Acp{RMHD} simulations that handle \ac{RT} properly alongside \acp{MHD} are
daunting for multiple reasons. Gas velocities at minidisk scales can be close
to the speed of light, so radiation propagation must be tracked by solving the
time-dependent form of the \ac{RT} equation. The complex disk structure could
mean that simple moment-based prescriptions of radiation anisotropy have to be
replaced by more accurate treatments. At temperatures \qty{\sim e5}{\kelvin}
typically encountered in \ac{MBH} minidisks, atomic opacity can be orders of
magnitude greater than free--free and free--bound opacities
\citep[e.g.,][]{2014ApJ...787....1H, 2022A&A...659A..87H, 2021MNRAS.508..453Z},
necessitating the use of opacities derived from detailed atomic calculations.
These additional requirements can make \acp{RMHD} simulations computationally
prohibitively expensive, tens or hundreds of times more so than their \acp{MHD}
counterparts \citep{2023ApJ...949..103W}. That said, \acp{RMHD} simulations
featuring detailed \ac{RT} is the required next step toward robust
observational predictions. Encouragingly, recent work on this front
demonstrates that \acp{RMHD} simulations of disks around single \acp{MBH}
\citep{2019ApJ...885..144J, 2023ApJ...945...57H} and circumbinary disks around
\acp{MBHB} \citep{2025arXiv250218584T} are feasible.

In this article, we report the first \acp{RMHD} simulation studying the
properties of \ac{MBHB} minidisks. We focus on \acp{MBHB} whose orbital
evolution is dominated by \ac{GW} emission. At the high-mass end, these
\acp{MBHB} are continuous sources in the \ac{PTA} band. At the low-mass end,
they are evolving in frequency into the \ac{LISA} band; in this sense, they are
the direct precursors to mergers \ac{LISA} will see. We find that our
\acp{RMHD} minidisk retains several properties characteristic of \acp{MHD}
minidisks in the literature, but at the same time differs from those minidisks
in structure and thermodynamics.

We describe our methods in \cref{sec:methods}, communicate our results in
\cref{sec:results}, discuss them in \cref{sec:discussion}, and wrap up in
\cref{sec:conclusions}.

\section{Methods}
\label{sec:methods}

\subsection{Problem setup and simulation strategy}

We consider \iac{MBHB} for which the total mass is
$M_{\su{tot}}=\qty{2e7}{\solarmass}$ and the orbital separation is
$a=100\,GM_{\su{tot}}/c^2$, where $G$ is the gravitational constant and $c$ is
the speed of light. The masses of the individual \acp{MBH} are
$M_1=M_2=\tfrac12M_{\su{tot}}$, with $M_1$ being the \ac{MBH} at the center of
the minidisk of interest and $M_2$ being the companion \ac{MBH}, Such
\iac{MBHB} can be detected by \ac{LISA} during late inspiral and merger, but
here we study a moment when the \ac{MBHB} is still outside the \ac{LISA} band.
Results of our simulations can be extrapolated to different $M_{\su{tot}}$,
such as those pertinent to \ac{PTA} detections; this we shall do in
\cref{sec:discussion}.

The orbital period of \iac{MBHB} with our chosen parameters is
\begin{equation}
t_{\su{orb}}=2\pi\biggl(\frac{a^3}{GM_{\su{tot}}}\biggr)^{1/2}
  \approx\qty{7.2}{\day}\times\frac{M_{\su{tot}}}{\qty{2e7}{\solarmass}},
\end{equation}
and its time to merger due to the emission of \acp{GW} is
\citep{1964PhRv..136.1224P}
\begin{equation}
t_{\su{gw}}=\frac{5c^5a^4}{256G^3M_1M_2M_{\su{tot}}}
  \approx\qty{24}{\year}\times\frac{M_{\su{tot}}}{\qty{2e7}{\solarmass}}.
\end{equation}
We follow the minidisk over several $t_{\su{orb}}$. As this is much shorter
than $t_{\su{gw}}$, we neglect the shrinking of the binary orbit.

We use the finite-volume Newtonian \acp{MHD} code Athena++
\citep{2020ApJS..249....4S}, coupled with the \ac{RT} module of
\citet{2021ApJS..253...49J}, to simulate the minidisk around one of the
\acp{MBH} in a circular equal-mass binary. This setup is advantageous because a
unified \acp{RMHD} simulation that tracks both the circumbinary disk and the
two \ac{MBH} minidisks, solving simultaneously the \acp{MHD} and time-dependent
\ac{RT} equations, is prohibitively expensive. Our simulation neglects the
interaction between the two minidisks, such as mass transfer due to sloshing
across the first Lagrangian point L1 and outflows, and energy transfer due to
mutual illumination. However, sloshing is unimportant at our chosen orbital
separation because of the high gravitational potential barrier at L1. Mass
transfer by outflows and mutual illumination are also minimal for aligned,
widely separated minidisks.

The simulation is set up in a corotating frame in which the binary is at rest
and the \ac{MBH} the minidisk orbits around is at the origin. The corotating
frame is reached from the inertial frame in two steps: first, a rotation
through an angle of $-\Omega t$, where $\Omega=(GM_{\su{tot}}/a^3)^{1/2}$ is
the binary orbital frequency and $t$ is the time, about an axis normal to the
binary orbital plane and going through the binary center of mass; second, a
translation along the line connecting the two \acp{MBH} from the binary center
of mass to the \ac{MBH} with the minidisk. We describe the corotating frame
with spherical coordinates $(r,\theta,\phi)$, such that $\theta=\tfrac12\pi$ is
the binary orbital plane and the companion \ac{MBH} is at
$(r,\theta,\phi)=(a,\tfrac12\pi,\pi)$.

To build the minidisk, we inject a gas stream from the boundary of the
computational domain in a manner similar to \citet{2017ApJ...835..199R}. This
setup idealizes one of the streams that originate from the inner edge of the
circumbinary disk and feed the minidisks. The stream is characterized in detail
in \cref{sec:stream}.

To further conserve computational resources, we divide our simulation into
stages \citep[see also][]{2017ApJ...843...58C} so as to give the minidisk ample
time to relax in \acp{MHD} before turning on expensive \acp{RMHD} evolution. We
describe how we implement the stages in \cref{sec:branches}. An additional
benefit of this approach is that we can isolate the impact of radiation on disk
properties by comparing the \acp{MHD} and \acp{RMHD} minidisks.

Even with these cost-saving measures, our simulation still requires millions of
core-hours to run. This limits our simulation duration to a few binary orbits
in \acp{MHD} plus a fraction of a binary orbit in \acp{RMHD}, too short for
following the slow inspiral of the \ac{MBHB} or computing the steady-state
torques on the gas. However, because the goal of this first \acp{RMHD} minidisk
simulation is to predict the fundamental observational character of a minidisk
shaped by radiation, it suffices that the emission properties of the inner
parts of the minidisk reach a steady state.

\subsection{\Aclp{RMHD}}

The equations of Newtonian ideal \acp{RMHD} in the corotating frame are
\begin{align}
\pd\rho t+\divg(\rho\vec v) &= 0, \\
\pd{}t(\rho\vec v)+\divg(\rho\vec v\vec v-\vec B\vec B+p^*\vtn I) &=
  \vec f+\vec g_1+\vec g_2+\vec S^{\su m}, \\
\pd Et+\divg[(E+p^*)\vec v-(\vec v\cdot\vec B)\vec B] &=
  \vec v\cdot(\vec f+\vec g_1+\vec g_2)+S^{\su e}, \\
\pd{\vec B}t-\curl(\vec v\vectimes\vec B) &= \vec0.
\end{align}
Here $\rho$, $\vec v$, and $p$ are the gas density, velocity, and pressure,
respectively, $\vec B$ is the magnetic field, $p^*=p+\tfrac12B^2$ and
$E=\tfrac12\rho v^2+p/(\gamma-1)+\tfrac12B^2$ are the total pressure and
energy, $\gamma=\tfrac53$ is the adiabatic index, and $\vtn I$ is the isotropic
rank-two tensor. The forces $\vec f$, $\vec g_1$, and $\vec g_2$ are defined in
the following paragraphs, and the momentum and energy source terms $\vec S^{\su
m}$ and $S^{\su e}$ due to radiation are given in the next section.

The fictitious force in the corotating frame is
\begin{equation}
\vec f=-2\rho\vec\Omega\vectimes\vec v
  -\rho\vec\Omega\vectimes[\vec\Omega\vectimes(\vec r+\vec a_1)],
\end{equation}
where $\vec\Omega$ is the vector normal to the binary orbital plane with
magnitude equal to the binary orbital frequency $\Omega$, and $\vec a_i$ are
the displacements in the corotating frame of the two \acp{MBH} from their
center of mass.

The gravitational forces due to the \ac{MBH} at the center of the minidisk and
the companion \ac{MBH} are $\vec g_1$ and $\vec g_2$, respectively. We
approximate the general-relativistic effects due to non-spinning \acp{MBH} with
the pseudo-Newtonian prescription of \citet{2013MNRAS.433.1930T}:
\begin{multline}
\vec g_i=\rho\bigg[
  -\frac{GM_i\vec r_i}{r_i^3}\biggl(1-\frac{2GM_i}{c^2r_i}\biggr)^2
  +\frac{2GM_i\vec v_i}{c^2r_i^3}\biggl(1-\frac{2GM_i}{c^2r_i}\biggr)
  ^{-1}(\vec r_i\cdot\vec v_i) \\
-\frac{3GM_i\vec r_i}{c^2r_i^5}\norm{\vec r_i\vectimes\vec v_i}^2\biggr],
\end{multline}
with
\begin{align}
\vec r_i &\eqdef \vec r+\vec a_1-\vec a_i, \\
\vec v_i &\eqdef \vec v+\vec\Omega\vectimes\vec r_i.
\end{align}
This velocity-dependent prescription imitates Schwarzschild spacetime more
accurately than the \citet{1980A&A....88...23P} potential. Note that
$\vec\Omega$, $\vec a_i$, $\vec f$, and $\vec g_i$ are all independent of time.

\subsection{\Acl{RT}}

In keeping with our Newtonian treatment of \acp{MHD}, we perform \ac{RT}
assuming flat spacetime. Our \ac{RT} scheme \citep{2018ApJ...854..110Z,
2019ApJ...880...67J, 2021ApJS..253...49J} discretizes the solid angle around
each cell into an angle grid of rays and solves the evolution equation of
intensity on each individual ray in lockstep with \acp{MHD}. The \ac{RT} scheme
takes into account the special-relativistic effect of Lorentz transformation
from the frame of one gas packet to that of another, but it ignores the
general-relativistic effects of gravitational redshift, light bending, and
Shapiro delay.

In the corotating frame in which the \acp{MHD} equations are written, the
time-dependent, frequency-integrated \ac{RT} equation is
\begin{equation}\label{eq:inertial RT}
\frac1c\pd{}tI(\uvec n)+\uvec n\cdot\grad I(\uvec n)=S(\uvec n),
\end{equation}
where $I(\uvec n)$ and $S(\uvec n)$ are the intensity and its source term,
respectively, in the direction $\uvec n$. In the frame instantaneously comoving
with a gas packet with velocity $\vec v$, the same two quantities are
\begin{align}
I_0(\uvec n_0) &= \delta^{-4}(\uvec n,\vec v)I(\uvec n), \\
S_0(\uvec n_0) &= \delta^{-3}(\uvec n,\vec v)S(\uvec n),
\end{align}
where
\begin{align}
\delta(\uvec n,\vec v) &= [\Gamma(1-\uvec n\cdot\vec v/c)]^{-1}, \\
\Gamma &= (1-v^2/c^2)^{-1/2}
\end{align}
are the Doppler and Lorentz factors, respectively, and
\begin{equation}
\uvec n_0=
  \frac{\uvec n-\Gamma(\vec v/c)[1-\Gamma(\uvec n\cdot\vec v/c)/(\Gamma+1)]}
  {\Gamma(1-\uvec n\cdot\vec v/c)}
\end{equation}
is $\uvec n$ boosted to the comoving frame. Therefore, the \ac{RT} equation can
also be written as
\begin{equation}\label{eq:mixed RT}
\frac1c\pd{}tI_0(\uvec n_0)+\uvec n\cdot\grad I_0(\uvec n_0)=
  \delta^{-1}(\uvec n,\vec v)S_0(\uvec n_0).
\end{equation}
The advective term is more easily handled in the corotating frame, but the
intensity source term is more intuitively expressed in the comoving frame.
Consequently, we evolve the \ac{RT} equation in an operator-split manner: the
advective step is performed in the corotating frame according to
\cref{eq:inertial RT}, the intensity is boosted into the comoving frame, the
intensity source step is applied in the comoving frame following \cref{eq:mixed
RT}, and finally the intensity is boosted back into the corotating frame.

The comoving intensity source term includes absorption, Thomson scattering, and
Compton scattering:
\begin{multline}
S_0(\uvec n_0)\eqdef\rho(\kappa_{\su{a,R}}+\kappa_{\su s})[J_0-I_0(\uvec n_0)]
  +\rho\kappa_{\su{a,P}}(B-J_0) \\
+\rho\kappa_{\su s}J_0\times4\kB(T-T_{\su C})/(m_{\particle e}c^2).
\end{multline}
Here $\kappa_{\su{a,R}}$, $\kappa_{\su{a,P}}$, and $\kappa_{\su s}$ are the
Rosseland-mean absorption opacity, Planck-mean absorption opacity, and
Rosseland-mean scattering opacity, respectively, $B=c\aSB T^4/(4\pi)$ is the
black-body mean intensity, $\aSB$ is the Stefan--Boltzmann constant, $\kB$ is
the Boltzmann constant, $T_{\su C}\eqdef[4\pi J_0/(c\aSB)]^{1/4}$ approximates
the Compton temperature, $m_{\particle e}$ is the electron mass, and
\begin{equation}
J_0=\frac1{4\pi}\int d^2\uvec n_0\,I_0(\uvec n_0).
\end{equation}
In calculating the gas temperature $T=(\mu/\kB)(p/\rho)$, we assume fully
ionized hydrogen and helium with respective mass fractions $X=0.7$ and $Y=1-X$;
this implies a mean particle mass of $\mu\approx0.6\,\mH$ and a mean mass per
electron of $\mu_{\particle e}\approx1.2\,\mH$, where $\mH$ is the hydrogen
mass. After the intensity source step, we calculate the change $\Delta I(\uvec
n)$ in intensity over the time step $\Delta t$, and add to the gas the amount
of energy and momentum the radiation lost as
\begin{align}
S^{\su e}\,\Delta t &\eqdef
  -\frac1c\int d^2\uvec n\,\Delta I(\uvec n), \\
\vec S^{\su m}\,\Delta t &\eqdef
  -\frac1{c^2}\int d^2\uvec n\,\uvec n\,\Delta I(\uvec n).
\end{align}

Our opacity prescription as described below is almost the same as
\citet{2019ApJ...885..144J}; the only difference is that we base our
prescription not on \textacsize{OPAL} tables \citep{1996ApJ...464..943I}, but
on the tables compiled by \citet{2021MNRAS.508..453Z}. These newer tables
provide, as functions of $(\rho,T)$, the Rosseland- and Planck-mean opacities
$(\tilde\kappa_{\su R},\tilde\kappa_{\su P})$ due to atoms, molecules, and dust
at solar metallicity. The tabulated opacities cannot be fed directly to a
simulation expecting $(\kappa_{\su{a,R}},\kappa_{\su{a,P}},\kappa_{\su s})$
because Thomson scattering is included in $\tilde\kappa_{\su R}$; instead, we
must separate out its contribution $\kappaT=\sigmaT/\mu_{\particle
e}\approx\qty{0.34}{\centi\meter\squared\per\gram}$, where $\sigmaT$ is the
Thomson cross section, as $\kappa_{\su s}$. If $\tilde\kappa_{\su
R}\ge\kappaT$, which is true for most $(\rho,T)$, we simply set
$(\kappa_{\su{a,R}},\kappa_{\su s})=(\tilde\kappa_{\su R}-\kappaT,\kappaT)$.
Otherwise, we use $T$ to decide whether the Rosseland-mean opacity is due to
Thomson scattering: One possibility is that the gas is either too cold to be
ionized, in which case we assign $(\kappa_{\su{a,R}},\kappa_{\su
s})=(\tilde\kappa_{\su R},0)$; the other possibility is that the gas is too hot
for atomic opacity to be significant, in which case we demand
$(\kappa_{\su{a,R}},\kappa_{\su s})=(0,\tilde\kappa_{\su R})$. Lastly, we
require that $\kappa_{\su{a,P}}=\max\{\tilde\kappa_{\su
P},\kappa_{\su{a,R}}\}$; note that $\tilde\kappa_{\su P}>\tilde\kappa_{\su R}$
generally for disk $(\rho,T)$ except for $T\gtrsim\qty{3e5}{\kelvin}$.

\subsection{Feeding the minidisk with a stream}
\label{sec:stream}

To represent accretion from the circumbinary disk, we inject a magnetized
stream toward the \ac{MBH} placed at the center of the simulation domain. The
injection point is $(r,\theta,\phi)=(r_{\su{inj}},\tfrac12\pi,0)$, where
$r_{\su{inj}}=\tfrac12a=50\,GM_{\su{tot}}/c^2$; this point lies along the line
joining the two \acp{MBH}, on the opposite side of the companion \ac{MBH}
sitting outside the simulation domain. The injection velocity is in the binary
orbital plane:
\begin{equation}
\vec v_{\su{inj}}=-0.1\,(GM_{\su{tot}}/a)^{1/2}(\uvec e_R+\uvec e_\phi)
\end{equation}
in the corotating frame; streams with injection velocities very different from
this tend not to make a tight loop around the \ac{MBH}. The pericenter distance
of the stream is $r\approx2.7\,GM_{\su{tot}}/c^2$, between the marginally bound
orbit $r=2\,GM_{\su{tot}}/c^2$ and the \ac{ISCO} $r=3\,GM_{\su{tot}}/c^2$ of
the \ac{MBH}, and it crosses itself at $r\approx6.4\,GM_{\su{tot}}/c^2$.

The stream is modeled as a cylinder. Stream properties are most conveniently
described in the rotated Cartesian coordinate system $(\tilde x,\tilde y,\tilde
z)$, whose positive $\tilde z$\nobreakdash-axis coincides with the stream axis
and points in the direction of stream motion, and whose $\tilde
x$\nobreakdash-axis lies in the binary orbital plane. The density of the stream
in this coordinate system is
\begin{equation}
\rho\propto\exp\biggl(-\frac{\tilde x^2+\tilde y^2}{w_{\su{inj}}^2}\biggr),
\end{equation}
where $w_{\su{inj}}=0.05\,a$ is the stream width. The normalization of the
equation is adjusted such that the mass injection rate is
\begin{equation}\label{eq:injection}
\dot M_{\su{inj}}=\frac{L_{\su{inj}}}{L_{\su E}}\frac{L_{\su E}}{\eta c^2},
\end{equation}
where we choose the Eddington ratio to be $L_{\su{inj}}/L_{\su E}=0.1$, $L_{\su
E}=4\pi GM_{\su{tot}}\mH c/\sigmaT$ being the Eddington luminosity, and the
accretion efficiency to be $\eta=0.1$. This implies a density of $\qty{\approx
e-11}{\gram\per\centi\meter\cubed}$ along the stream axis. We use a
time-independent $\dot M_{\su{inj}}$ so we can form a time-steady minidisk,
allowing us to disentangle observational time-variation due to minidisk
nonaxisymmetry and orbital motion from other effects. When simulating the
minidisk in \acp{MHD}, we set the gas pressure of the stream to $p=\rho
c_{\su{s,inj}}^2$, where $c_{\su{s,inj}}=(1/30)(GM_{\su{tot}}/a)^{1/2}$. When
simulating the minidisk in \acp{RMHD}, we set the sum of gas and radiation
pressures $p+\tfrac43\pi J_0$ to the same, requiring additionally that gas and
radiation be in thermal equilibrium; this produces a radiation-dominated stream
with $T\sim\qty{8e4}{\kelvin}$.

The magnetic field in the stream is derived from the magnetic potential
\begin{multline}
\vec A_{\su{inj}}\propto
  \sin\biggl(\pi\frac{\tilde z-v_{\su{inj}}t}{w_{\su{inj}}}\biggr)\times{} \\
\biggl[1+\cos\biggl(
  \pi\min\biggl\{1,\frac{(\tilde x^2+\tilde y^2)^{1/2}}{w_{\su{inj}}}\biggr\}
  \biggr)\biggr]\,\uvec e_{\tilde x}.
\end{multline}
The magnetic potential paints onto the stream frozen-in magnetic field loops
that are perpendicular to the binary orbital plane. The loops have alternating
circulation, minimizing the bias we introduce to the magnetic field direction
in the minidisk. Perpendicular loops are preferred over coplanar ones because
horizontal magnetic fields cannot create vertical magnetic fields through
differential rotation alone. The magnetic field is normalized such that the
volume-integrated plasma beta, defined as the magnetic pressure integrated over
the stream divided by the sum of gas and radiation pressures similarly
integrated, is unity. This plasma beta is representative of streams in
circumbinary disk simulations \citep{2025arXiv250218584T}. As a practical
matter, we find that injecting a more strongly magnetized stream helps with
resolving \ac{MHD} turbulence in the minidisk later.

Our prescription for introducing matter to the vicinity of the \ac{MBH} models
only crudely the flow from the circumbinary disk to the minidisk. In reality,
gas with a range of energies, angular momenta, and magnetization approaches the
minidisk from different directions, the trajectories and properties of the
incoming gas vary over time, and some of the gas may fail to join the minidisk
and be flung back out to the circumbinary disk
\citep[e.g.,][]{2022ApJ...932...24T}. Our simplified scheme is justified
insofar as material accreted to the minidisk is rapidly mixed, so that our
stream reflects the time-averaged properties of the more complicated flow.

Our choice of stream parameters leads to the formation of a compact minidisk,
smaller than the effective radius of the Roche lobe of the \ac{MBH}. This is
largely a consequence of our finite simulation time; if we were able to
continue the simulation indefinitely, the minidisk should expand to fill the
effective radius. Our compact minidisk does not exhibit spiral features because
it is not subject to strong tidal forces from the companion \ac{MBH}. However,
we do not expect the compactness of the minidisk to change our observational
predictions qualitatively, because minidisk emission, much of it thermal, is
dominated by the innermost parts of the minidisk.

\subsection{Splitting the simulation into \acp{MHD} and \acp{RMHD}}
\label{sec:branches}

Evolving the \acp{MHD} equations together with the time-dependent \ac{RT}
equations on multiple rays is \num{\sim10} times slower than evolving just the
\acp{MHD} equations. Therefore, we partition our simulation into two stages
\citep[see also][]{2017ApJ...843...58C}. The first stage, performed entirely in
\acp{MHD}, follows minidisk formation. The second stage, consisting of
concurrent \acp{MHD} and \acp{RMHD} branches, allows us to compare our minidisk
with \acp{MHD} simulations in the literature and to examine the impact of
radiation by contrasting the \acp{MHD} and \acp{RMHD} branches.

In the first stage, we follow gas and magnetic field buildup in the minidisk in
\acp{MHD}. We employ a locally isothermal equation of state: for every time
step, we evolve \acp{MHD} under an adiabatic equation of state with
$\gamma=\tfrac53$, and then we set the gas pressure to
$p=(1/30)^2(GM_1\rho/r)$. This locally isothermal equation of state imitates
rapid cooling and encourages minidisk formation. The simulation is run in this
stage for $5.4\,t_{\su{orb}}$, or equivalently, \num{\sim260} orbits at
$r\sim6\,GM_{\su{tot}}/c^2$. The minidisk density distribution does not change
qualitatively after $t\sim t_{\su{orb}}$.

In the second stage, the simulation splits into \acp{MHD} and \acp{RMHD}
branches. The \acp{MHD} branch is a straightforward extension of the first
stage. The \acp{RMHD} branch requires us to first replace the gas pressure
everywhere by a combination of gas and radiation pressures under thermal
equilibrium: given the gas temperature $T_{\su{old}}$ at the end of the first
stage, we solve the equation $\rho\kB T_{\su{old}}/\mu=\tfrac13\aSB
T_{\su{new}}^4+\rho\kB T_{\su{new}}/\mu$ for the gas and radiation temperature
$T_{\su{new}}$ at the start of the \acp{RMHD} branch. We then continue the
simulation in full \acp{RMHD} using an adiabatic equation of state with
$\gamma=\tfrac53$. The second stage is terminated after $0.46\,t_{\su{orb}}$,
or \num{\sim22} orbits at $r\sim6\,GM_{\su{tot}}/c^2$, at which point the
minidisk density distribution has adjusted to the different force distribution
in the presence of radiation and the minidisk luminosity has reached steady
state.

\subsection{Simulation domain and boundary conditions}

The simulation domain stretches from the stream injection point down to the
photon orbit of the \ac{MBH} and covers the full solid angle around the
\ac{MBH}. The $r$\nobreakdash-, $\theta$\nobreakdash-, and
$\phi$\nobreakdash-coordinates range over $[1.5,50]$ in units of
$GM_{\su{tot}}/c^2$, $[0,\pi]$, and $[0,2\pi]$, respectively. The spatial grid
is uniform in the polar and azimuthal directions but logarithmic in the radial
direction. On top of that, we impose five levels of static mesh refinement on
the region $r\le10\,GM_{\su{tot}}/c^2$ and $\abs{\theta-\tfrac12\pi}\le0.1$,
achieving an effective resolution of $1024\times768\times768$ and a cell size
of $(\Delta r/r,\Delta\theta,\Delta\phi)\approx(3,4,8)\times\num{e-3}$ at the
highest resolution. Thanks to refinement, more than half of the radial cells
are at $r\lesssim5\,GM_{\su{tot}}/c^2$. Intensity is discretized on an angle
grid consisting of 80 rays. The angle grid is identical for all cells and does
not rotate with the coordinate basis vectors.

The radial direction uses diode boundary conditions: if a cell in the ghost
zone is occupied by the injected stream, we set its properties to those of the
stream; otherwise, we copy all variables from the last physical cell, and we
zero the radial velocity component and all magnetic field components if the
velocity points inward. The polar direction uses polar boundary conditions that
allow gas and radiation to pass over the pole. The azimuthal direction uses
periodic boundary conditions.

\section{Results}
\label{sec:results}

\subsection{Minidisk structure}

\begin{figure}
\includegraphics{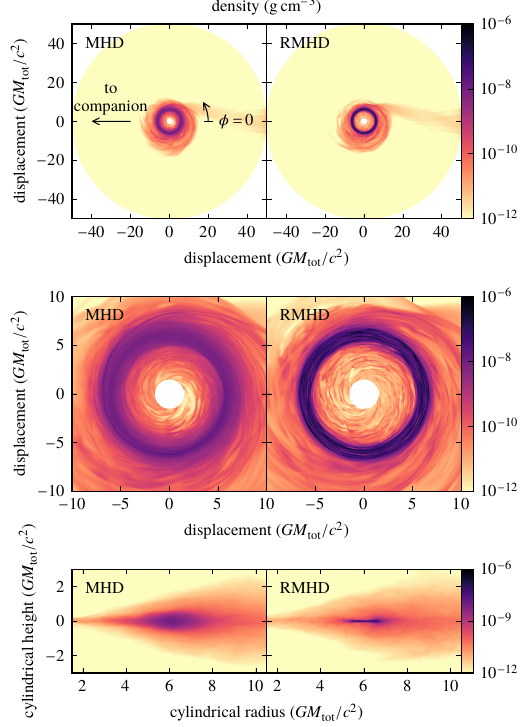}
\caption{Density of the minidisk at the end of the \acp{MHD} branch in the left
column and the \acp{RMHD} branch in the right column. The top row shows
midplane slices, the middle row shows zoomed-in versions of the same, and the
bottom row shows zoomed-in azimuthal averages. In all panels, one \ac{MBH} of
the binary is at the origin, the other is to the left outside the simulation
domain, and the stream is injected from the right.}
\label{fig:density}
\end{figure}

\Cref{fig:density} portrays the minidisk at the end of the \acp{MHD} and
\acp{RMHD} branches. The stream mimicking gas supply from the circumbinary disk
enters from the right. Even though the stream has comparable initial radial and
azimuthal velocities in the rotating frame, because these velocities are so
small, the stream appears almost radial. After looping around the \ac{MBH}, the
stream self-intersects and forms a minidisk. The minidisk consists of a
circular, geometrically thin ring of dense gas at $r\sim6\,GM_{\su{tot}}/c^2$
as well as a diffuse envelope that extends radially to
$r\sim12\,GM_{\su{tot}}/c^2$ and vertically above and below the stream plane.
Gas in both the ring and the envelope feeds the accretion flow toward the
\ac{MBH} at $r\lesssim5\,GM_{\su{tot}}/c^2$. All these structures are nestled
well within the Roche lobe of the \ac{MBH}, which has an effective radius
$\mathrelp\approx38\,GM_{\su{tot}}/c^2$ \citep{1959cbs..book.....K,
1983ApJ...268..368E}.

The stream feeding the minidisk does not directly collide with the ring;
rather, it shocks against the geometrically thicker and less dense envelope
surrounding the ring, intermixing into the envelope and eventually the ring.
The geometrical thinness of the ring is the result of rapid cooling: in the
\acp{MHD} case, it is emulated through the use of a locally isothermal equation
of state; in the \acp{RMHD} case, it is due to the actual escape of radiation
from the minidisk. The aspect ratio of the ring, defined as the mass-weighted
average of $\abs{\theta-\tfrac12\pi}$, is \num{\sim3e-2} in the \acp{MHD}
minidisk and \num{\sim4e-3} in the \acp{RMHD} minidisk. The former agrees with
our locally isothermal equation of state given in \cref{sec:branches}, and the
latter is at our resolution limit.

The ring, envelope, and accretion flow are persistent minidisk components.
Moreover, the compact minidisk does not appear strongly perturbed by the tidal
forces of the companion \ac{MBH}. These properties contrast with the transience
and strong tidal distortion characteristic of minidisks in tighter binaries
with orbital separations $\mathrelp\lesssim20\,GM_{\su{tot}}/c^2$
\citep[e.g.,][]{2017ApJ...838...42B, 2023MNRAS.520..392B}. In our simulation
with a limited duration, most of the gas supplied by the stream stays in the
persistent structures, and only a fraction is accreted onto the \ac{MBH} over
the course of either the \acp{MHD} or the \acp{RMHD} branch.

\subsection{Accretion from the minidisk onto the \ac{MBH}}

\begin{figure}
\includegraphics{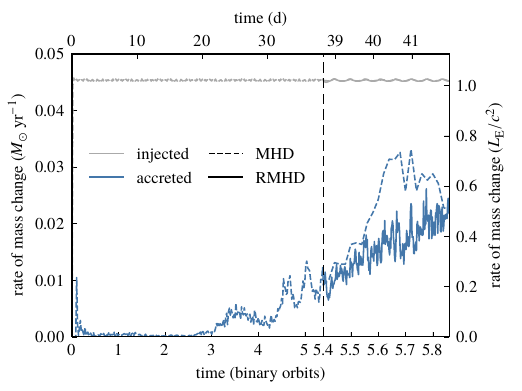}
\caption{Rates at which mass enters and exits the simulation domain. The left
and right panels are for the first and second stages respectively; the right
panel has a more expanded abscissa than the left panel. Colors differentiate
rates due to injection through the stream and accretion onto the \ac{MBH}. Line
styles indicate whether the simulation is in \acp{MHD} or \acp{RMHD}; the first
stage is only in \acp{MHD}, whereas the second stage has \acp{MHD} and
\acp{RMHD} branches.}
\label{fig:mass}
\end{figure}

\Cref{fig:mass} compares the rate of mass accretion onto the \ac{MBH} with the
rate of mass injection through the stream. As explained in \cref{sec:stream},
the injection rate is kept constant in this first \acp{RMHD} simulation of the
minidisk in order to distinguish the effects of time-steady feeding from
time-variable feeding.

The accretion rate is measured to be below the injection rate throughout the
simulation. This means the minidisk as a whole is not in inflow equilibrium;
rather, the ring accumulates mass in the net, even after the simulation has run
for \num{\approx282} orbits at $r\sim6\,GM_{\su{tot}}/c^2$, the radius of the
ring. Nevertheless, because \cref{sec:quality} shows the inner parts of the
minidisk to have sufficiently high quality factors, those parts are likely in
inflow equilibrium.

Even though the entire minidisk is not in inflow equilibrium, it is in
radiative equilibrium because, as we shall see later in \cref{sec:radiation},
the luminosity from the minidisk is roughly constant over time at the end of
the \acp{RMHD} branch. Radiative equilibrium is likely a more important concern
if the focus of our study is the emission properties of the minidisk.

\subsection{Pressure contributions in the minidisk}

\begin{figure}
\includegraphics{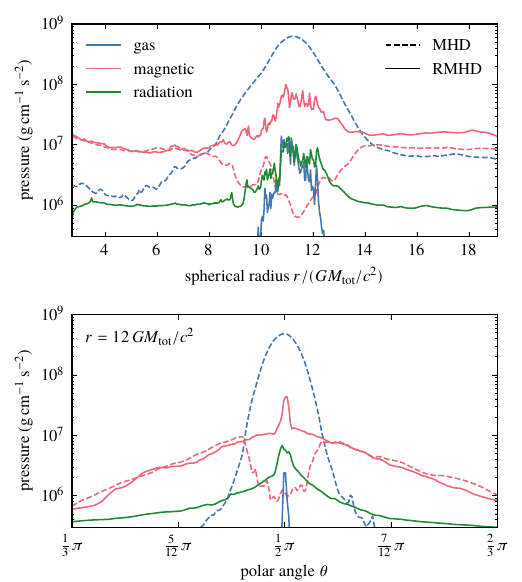}
\caption{Mass-weighted profiles of pressure contributions in the minidisk over
spherical radius and polar angle at the end of the \acp{MHD} and \acp{RMHD}
branches. Colors differentiate pressure contributions. Line styles indicate
whether the simulation is \acp{MHD} or \acp{RMHD}.}
\label{fig:pressure profile}
\end{figure}

The top and bottom panels of \cref{fig:pressure profile} plot, respectively,
the mass-weighted profiles of gas, magnetic, and radiation pressures in the
minidisk over $r$ and $\theta$ at the end of the \acp{MHD} and \acp{RMHD}
branches:
\begin{align}
\mean{\symcal P}_{\theta\phi;\rho} &=
  \iint\sin\theta\,d\theta\,d\phi\,\rho\symcal P\bigg/
  \iint\sin\theta\,d\theta\,d\phi\,\rho, \\
\mean{\symcal P}_{\phi;\rho} &= \evalb[\bigg]{
  \int d\phi\,\rho\symcal P\bigg/
  \int d\phi\,\rho
  }_{r=6\,GM_{\su{tot}}/c^2},
\end{align}
where $\symcal P\eqdef p$, $\symcal P\eqdef\tfrac12B^2$, and $\symcal
P\eqdef4\pi J_0/(3c)$ for gas, magnetic, and radiation pressures, respectively.
Mass-weighting is used here to pick out high-density regions.

The dominant pressure contribution differs between the \acp{MHD} and \acp{RMHD}
minidisks. In the \acp{MHD} minidisk, gas pressure is up to \numrange{\sim2}{3}
orders of magnitude higher than magnetic pressure. The ratio of gas to magnetic
pressure is greatest at $r\sim6\,GM_{\su{tot}}/c^2$ because gas from the stream
accumulates as a dense ring at that radius. In the \acp{RMHD} minidisk,
magnetic pressure exceeds radiation pressure by a factor of
\numrange{\sim2}{10} and gas pressure is negligible. The ratio of radiation to
magnetic pressure is greatest in the ring because the dense gas expels the
buoyant magnetic field.

The pressure contribution responsible for supporting the minidisk against
vertical gravity also differs between the \acp{MHD} and \acp{RMHD} minidisks.
In the \acp{MHD} minidisk, support is mainly due to gas pressure near the
midplane, switching to a mixture of gas and magnetic pressures further from the
midplane. In the \acp{RMHD} minidisk, magnetic and radiation pressures serve
the same purpose. The amount of pressure support in the \acp{MHD} minidisk is
set by the locally isothermal equation of state. After the transition to
\acp{RMHD}, the minidisk loses much of that pressure support as radiation
escapes the minidisk. This is evidenced by the shallower pressure gradients in
and smaller geometrical thickness of the ring in the \acp{RMHD} minidisk.

The horizontal profiles of gas pressure in the \acp{MHD} minidisk and radiation
pressure in the \acp{RMHD} minidisk are rather similar in shape. However, the
latter profile is lower by \num{\sim1} order of magnitude, and the difference
is greater in the ring where the gas is denser. Radiation pressure in
\acp{RMHD} behaves differently from gas pressure in \acp{MHD} because radiation
is not strictly tied to higher-density gas where it is produced and tends to
escape to regions with lower-density gas.

\subsection{Radiation generation and propagation in the minidisk}
\label{sec:radiation}

The \acp{RMHD} minidisk is in a state of radiative equilibrium. Most of the
radiation is produced by dissipation internal to the minidisk eventually
escapes to infinity. A negligible amount of energy is delivered by the stream
or captured by the \ac{MBH}.

\begin{figure}
\includegraphics{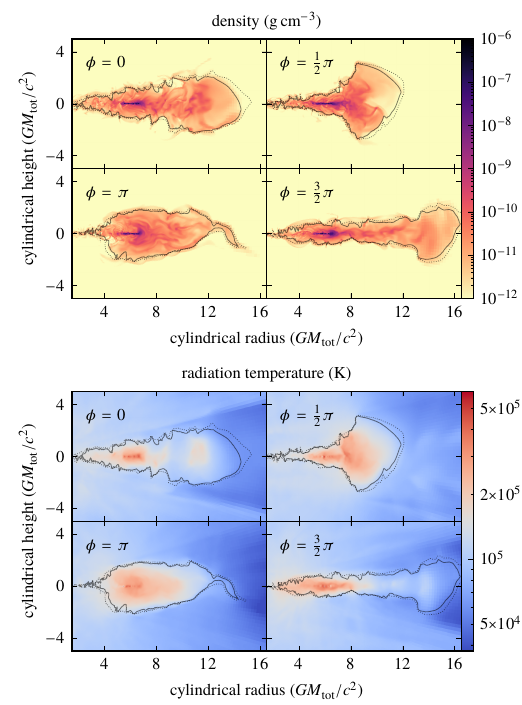}
\caption{Density and radiation temperature of the minidisk at the end of the
\acp{RMHD} branch on various poloidal slices. The solid and dotted contours are
the effective photosphere and scattering photosphere, respectively.}
\label{fig:radiation}
\end{figure}

The top panel of \cref{fig:radiation} depicts again a poloidal view of the
minidisk density, but on individual poloidal slices instead of azimuthally
averaged as in \cref{fig:density}. It is apparent that the minidisk is highly
nonaxisymmetric. The radial extent of the minidisk and its vertical thickness
both vary with azimuth.

The solid contours trace the effective photosphere. The location of the
effective photosphere from above is defined to be the polar angle $\theta_{\su
a}(r,\phi)$ satisfying
\begin{equation}
\int_0^{\theta_{\su a}} r\,d\theta\,\rho\kappa\eff\eqdef1,
\end{equation}
and the location of the same from below is defined to be $\theta_{\su
b}(r,\phi)$ satisfying
\begin{equation}
\int_{\theta_{\su b}}^\pi r\,d\theta\,\rho\kappa\eff\eqdef1,
\end{equation}
where
\begin{equation}
\kappa\eff=[\kappa_{\su{a,R}}(\kappa_{\su{a,R}}+\kappa_{\su s})]^{1/2}
\end{equation}
is the effective opacity. The effective photosphere, $\theta_{\su a}$, and
$\theta_{\su b}$ are undefined at some $(r,\phi)$ if the gas there is so
optically thin that
\begin{equation}
\int_0^\pi r\,d\theta\,\rho\kappa\eff<2.
\end{equation}
The effective photosphere thus defined tells us approximately where radiation
is last thermalized before it escapes to infinity. The dotted contours follow
the scattering photosphere, defined analogously as the effective photosphere
but using $\kappa_{\su s}$ instead of $\kappa\eff$ in the integrands. Both
photospheres inherit the nonaxisymmetry of the gas. The proximity of the two
photospheres to each other is because the $\kappa_{\su{a,R}}$ is generally not
much greater than $\kappa_{\su s}$. Lastly, the minidisk is not optically thick
all the way down to the \ac{MBH} at all azimuths in effective opacity terms.

\begin{figure}
\includegraphics{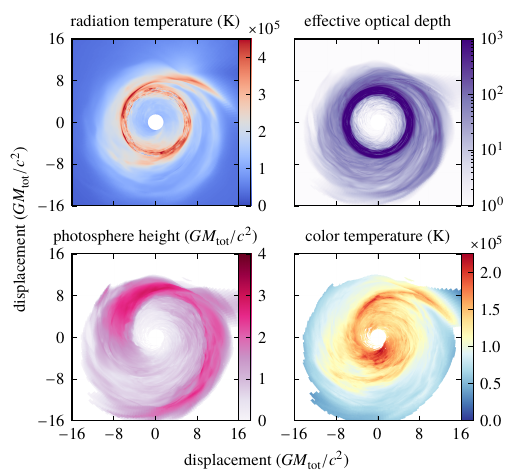}
\caption{Radiation properties of the minidisk at the end of the \acp{RMHD}
branch. \textit{Top-left panel:} Radiation temperature in the midplane.
\textit{Top-right panel:} Effective optical depth in the vertical direction.
\textit{Bottom-left panel:} Height of the effective photosphere above the
midplane. \textit{Bottom-right panel:} Radiation temperature on the effective
photosphere. For the bottom panels, quantities shown are the averages for the
effective photospheres from above and below, and empty regions have such low
optical depths that the effective photosphere is undefined.}
\label{fig:photosphere}
\end{figure}

The bottom panel of \cref{fig:radiation} presents on the same poloidal slices
the radiation energy $4\pi J_0/c$, expressed in terms of the radiation
temperature $T_{\su{rad}}=[4\pi J_0/(c\aSB)]^{1/4}$. The top-left panel of
\cref{fig:photosphere} shows the radiation temperature in the midplane.
Radiation is distributed nonaxisymmetrically like gas. Regions of denser gas
tend to have more intense radiation; however, because radiation is not strictly
bound to the gas, distributions of gas and radiation are not identical.
Notably, the suppression of radiation within a wedge-shaped region above and
below the midplane at large radii is indicative of shadowing of emission from
the hotter inner minidisk by the colder outer minidisk. Furthermore, certain
features of the radiation distribution can be explained in physical terms: for
example, the concentration along the $r\approx6\,GM_{\su{tot}}$ circle is due
to the presence of the dense ring, whereas the arc running across the top half
of the midplane panel follows the spiral shock formed when the stream collides
with the less dense envelope around the dense ring. The temperature along the
spiral shock is \SIrange{\sim3e5}{4e5}{\kelvin}, a result of efficient
thermalization by dense gas with high absorption opacity. Our simulation does
not produce post-shock temperatures of \SI{\sim e9}{\kelvin} anticipated by
\citet{2014ApJ...785..115R}, meaning that, at least in our setup, the
hypothesized \SI{\sim100}{\kilo\electronvolt} emission from the contact hotspot
may be absent.

The top-right panel of \cref{fig:photosphere} displays the effective optical
depth:
\begin{equation}
\tau\eff=\frac12\int r\,d\theta\,\rho\kappa\eff.
\end{equation}
In agreement with the effective photosphere contours in \cref{fig:radiation},
the minidisk is optically thick over most of its surface area. Considering its
optical thickness, henceforth we consider only the thermal radiation from the
minidisk, whose spectrum we approximate as a sum of black bodies.

The bottom-left panel of \cref{fig:photosphere} demonstrates that the height of
the effective photosphere above the midplane,
\begin{equation}
h\eff\eqdef\frac12r(\theta_{\su b}-\theta_{\su a}),
\end{equation}
is strongly modulated over the minidisk. The nonaxisymmetric minidisk shape is
approximately steady over time because it is the result of vertical gas motion
excited by the stream impacting the minidisk on one side.

The nonaxisymmetric radiation distribution and minidisk height mean that the
radiation temperature on the effective photosphere, or the color temperature of
the black-body emission, varies from point to point. The bottom-right panel of
\cref{fig:photosphere} plots the color temperature, computed as
$\tfrac12(T_{\su{rad}}(r,\theta_{\su a},\phi)+T_{\su{rad}}(r,\theta_{\su
b},\phi))$. The color temperature is \qtyrange{\sim1e5}{2e5}{\kelvin},
generally decreasing with distance from the \ac{MBH}. It reaches a maximum
close to the inner edge of the minidisk, at $r\sim2.5\,GM_{\su{tot}}/c^2$ and
$\phi\sim\tfrac32\pi$, thanks to a slightly higher midplane radiation
temperature and a smaller $h\eff$ and $\tau\eff$ there. By contrast, the
midplane radiation temperature reaches a maximum along the dense ring at
$r\sim6\,GM_{\su{tot}}/c^2$. The color temperature is also high along an arc
stretching across the top half of the panel, reflecting easier radiation
leakage from a region where the effective photosphere is almost vertical.

We reiterate that simple temperatures and optical depth estimates of the
minidisk cannot be relied upon to predict its luminous output. Rather, we need
to perform \ac{RT} to accurately follow radiation propagation through the
complex gas distribution, particularly in the vicinity of the effective
photosphere. This is where our scheme excels compared to others such as
flux-limited diffusion.

\subsection{Emission from a single minidisk in the equal-mass \ac{MBHB}}

\begin{figure}
\includegraphics{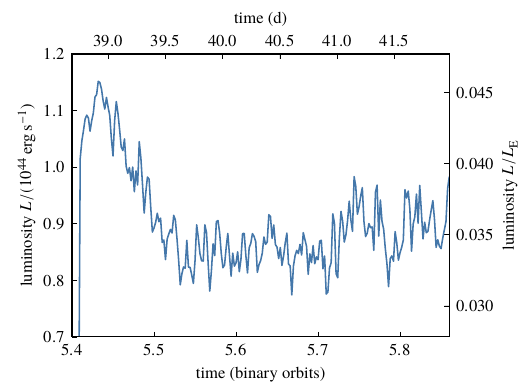}
\caption{Luminosity of the \acp{RMHD} minidisk leaving the simulation domain as
a function of time.}
\label{fig:luminosity}
\end{figure}

\Cref{fig:luminosity} traces the luminosity over time of the \acp{RMHD}
minidisk, quantified by the rate at which radiation leaves the simulation
domain. The initial transient relaxation phase lasts from $5.4\,t_{\su{orb}}$,
at the start of the \acp{RMHD} branch, to $\mathrelp\approx5.55\,t_{\su{orb}}$.
The minidisk luminosity then stabilizes to $\mathrelp\approx0.035\,L_{\su E}$,
or \qty{\approx7}{\percent} of the Eddington luminosity of the \ac{MBH} at the
center of the minidisk. This luminosity is in the expected range because the
stream injects gas at the rate $\dot M_{\su{inj}}=L_{\su E}/c^2$ according to
\cref{eq:injection}, the gas is injected almost from rest at
$r=50\,GM_{\su{tot}}/c^2$ with a specific binding energy of $0.01\,c^2$, and
most of the gas ends up in a circular orbit at $r\sim6\,GM_{\su{tot}}/c^2$ with
a specific binding energy of $\mathrelp\sim0.04\,c^2$.

\begin{figure}
\includegraphics{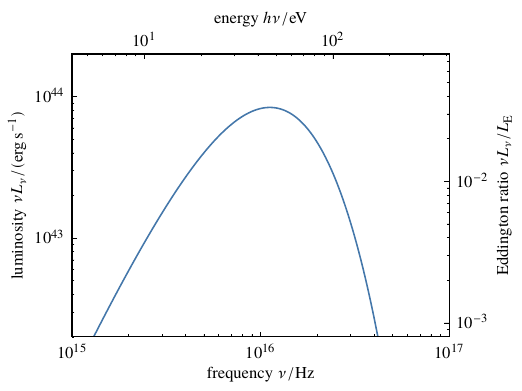}
\caption{Spectrum of the black-body emission from the minidisk at the end of
the \acp{RMHD} branch. Here $h$ is the Planck constant.}
\label{fig:spectrum}
\end{figure}

In \cref{sec:radiation}, we looked at the color temperature as a function of
location on the minidisk, that is, the radiation temperature
$T_{\su{rad}}(r,\theta_{\su{a,b}},\phi)$ as measured on the photosphere. Based
on this, here we estimate the spectrum of the entire minidisk by adding up the
black-body emission from all parts of the effective photosphere:
\begin{multline}
L_\nu(\nu)\eqdef
  \int dr\,r\sin\theta_{\su a}\,d\phi\,
  \pi B(\nu;T_{\su{rad}}(r,\theta_{\su a},\phi)) \\
+\int dr\,r\sin\theta_{\su b}\,d\phi\,
  \pi B(\nu;T_{\su{rad}}(r,\theta_{\su b},\phi)).
\end{multline}
The result is plotted in \cref{fig:spectrum}. The thermal spectrum peaks in the
extreme \ac{UV}, as expected for accretion disks around \acp{MBHB}.

\begin{figure}
\includegraphics{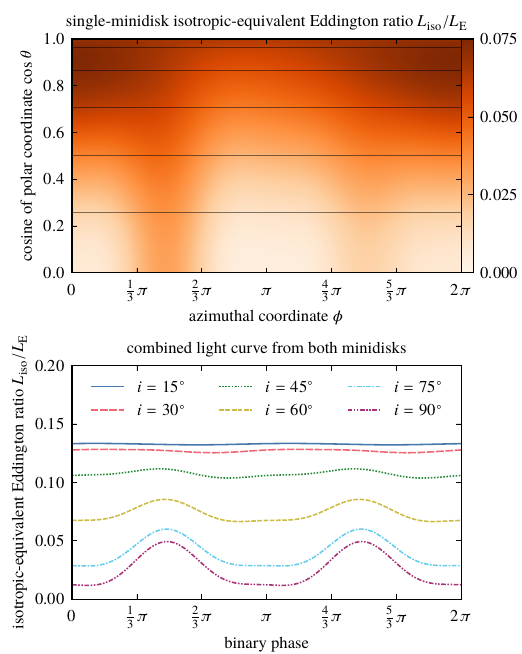}
\caption{\textit{Top panel:} Angular distribution of the radial radiative flux
from one minidisk in the corotating frame, measured at the end of the
\acp{RMHD} branch and at the outer boundary of the simulation domain, and
expressed in terms an isotropic-equivalent Eddington ratio. We limit $\theta$
to above the midplane because what is shown is the average of above- and
below-midplane values. \textit{Bottom panel:} Combined radial radiative flux
from both minidisks in the inertial frame as a function of time according to
observers at various inclinations $i$. Calculations are based on
single-minidisk results in the top panel and results are presented as
isotropic-equivalent Eddington ratios. The chosen inclinations are marked in
the top panel by horizontal lines.}
\label{fig:radiative flux}
\end{figure}

\Cref{fig:luminosity,fig:spectrum} are concerned with emission integrated over
viewing angle around the minidisk. However, emission can be highly anisotropic
because the effective photosphere is nonaxisymmetric. The top panel of
\cref{fig:radiative flux} illustrates the viewing-angle dependence of the
radial radiative flux $\uvec r\cdot4\pi\vec H$ in the frame corotating with the
\ac{MBHB}. To smooth out numerical artifacts due to angle-grid discreteness, we
expand the angular distribution of radiative flux in spherical harmonics
$Y_l^m(\theta,\phi)$ and retain only terms with $0\le l\le4$. The smoothed
radiative flux is quoted as an isotropic-equivalent Eddington ratio
$L_{\su{iso}}/L_{\su E}=4\pi r^2\uvec r\cdot4\pi\vec H/L_{\su E}$ for
comparison with observations.

Like the shape of the effective photosphere, the angular distribution of the
radiative flux is also roughly time-steady. At fixed $\phi$, the radiative flux
is stronger around the pole than near the midplane. At fixed $\theta$, the
radiative flux exhibits modulation over $\phi$ by \qty{\sim5}{\percent} for
$\theta$ around the pole and by a factor of \num{\sim7} for $\theta$ near the
midplane. This modulation is due to minidisk nonaxisymmetry and not special
relativistic beaming because the latter effect is not strongly
azimuth-dependent.

\subsection{Emission from both minidisks in the equal-mass \ac{MBHB}}

The nonaxisymmetric photosphere and anisotropic angular distribution of
radiative flux appear time-steady in the frame corotating with the \ac{MBHB}.
Even so, the radiative flux from both minidisks as perceived by a distant
observer in the inertial frame would still vary over time. Consider the
simplest case in which we approximate the two minidisks as identical and
aligned with the binary orbital plane, but rotated by $\pi$ with respect to one
another. At any moment in time, an observer at an inclination $i$ to the binary
orbital plane registers from the two minidisks an isotropic-equivalent
luminosity of $L_{\su{iso}}(i,\phi)+L_{\su{iso}}(i,\phi+\pi)$ for some $\phi$.
The viewing angle to the two minidisks change as the two minidisks go around
each other, so we have $\phi=2\pi t/t_{\su{orb}}$. Consequently, the light
curve the observer sees is periodic with a period of $\tfrac12t_{\su{orb}}$,
which is \qty{\sim3.6}{\day} for our choice of \ac{MBHB} parameters.

We showcase light curves for several $i$ in the bottom panel of
\cref{fig:radiative flux}. The directions of concentrated radiative flux of the
two minidisks sweep around like a lighthouse, producing the rise and fall in
the light curves. For a face-on observer, the radiative flux is strong and
time-steady. For an edge-on observer, the radiative flux is weak in a
time-averaged sense but its modulation amplitude is by a factor of a few.
Notably, the phase of the light variation is tied to the binary orbital phase,
which is in turn tied to the phase of the gravitational waves the \ac{MBHB}
emits.

\section{Discussion}
\label{sec:discussion}

\subsection{General observational implications}

Our simulation examines the emission properties of minidisks in an equal-mass
\ac{MBHB} with total mass $M_{\su{tot}}=\qty{2e7}{\solarmass}$ and orbital
separation $100\,GM_{\su{tot}}/c^2$. We found that minidisk nonaxisymmetry can
contribute to brightness variation with a period equal to half the binary
orbital period, and an amplitude equal to a factor of \num{\sim7} for edge-on
inclinations or a few for moderate inclinations. The large amplitude of this
variation potentially makes it stand out against the red-noise variability of
regular \acp{AGN} \citep[e.g.,][]{1996ApJ...470..364E, 2010ApJ...721.1014M,
2011ApJ...743L..12M} and against other nuclear transients such as tidal
disruption events. Its clearly interpretable periodicity is useful for
measuring the total mass and orbital separation of the \ac{MBHB}.

Special and general relativistic effects omitted from the simulation may modify
the observed brightness variation. As shown in \cref{fig:radiative flux}, the
minidisk concentrates its radiative flux toward $\phi\sim\tfrac12\pi$ in its
corotating frame. This direction points toward the observer when the minidisk
moves toward the observer as a result of the binary orbit; therefore, Doppler
boosting \citep{2010ApJ...715.1117B, 2015Natur.525..351D} can enhance the
already high level of brightness due to minidisk nonaxisymmetry. Additionally,
lensing of one minidisk by its companion \ac{MBH} can cause its observed
brightness to rise by a factor of a few \citep{2018MNRAS.474.2975D,
2022PhRvD.105j3010D, 2022PhRvL.128s1101D}. Because lensing requires the two
\acp{MBH} to be aligned along the sightline, it introduces additional
brightness peaks halfway in time between those due to minidisk nonaxisymmetry.
How exactly boosting and lensing alter the time-variation of brightness depends
on the angular distribution of radiative flux from each minidisk, the
inclination of the observer, the orientations of the minidisks, and so on. The
difference between these two relativistic effects and minidisk nonaxisymmetry
is that relativistic effects are strong only for tight, nearly edge-on
\acp{MBHB}, whereas minidisk nonaxisymmetry produces brightness variation over
a wider range of orbital separations and inclinations.

Quasiperiodic modulation in the accretion rate from the circumbinary disk of
equal-mass \acp{MBHB} may also lead to brightness variation with periods
comparable to the binary orbital period
\citetext{\citealp[e.g.,][]{2008ApJ...672...83M}\multicitedelim see
\citealp{2022LRR....25....3B} for a review}. The effect is most prominent for
\acp{MBHB} with orbital separations $\mathrelp\lesssim20\,GM_{\su{tot}}/c^2$,
whose compact, short-lived minidisks have inflow times shorter than the
timescale on which the accretion rate from the circumbinary disk varies
\citep[e.g.,][]{2022ApJ...928..187C, 2022ApJ...928..137G, 2023MNRAS.520..392B}.
Our minidisk, by contrast, is persistent because we consider orbital
separations $\mathrelp\sim100\,GM_{\su{tot}}/c^2$. Therefore, we expect the
effect of time-varying mass supply rate to be smoothed out.

We have so far focused on thermal emission from the minidisks. In contrast to
the simple picture in which the \acp{MBH} are surrounded either by
geometrically thin, optically thick disks characterized by thermal emission or
by geometrically thick, optically thin clouds characterized by non-thermal
emission, our minidisk is actually a combination of the two. At
$r\le16\,GM_{\su{tot}}/c^2$ where the minidisk dominates the stream, \num{\sim
e-2} of the gas by mass lies above the effective photosphere, and \num{\sim
e-2} of this gas has temperatures $\SI{\gtrsim0.1}{\kilo\electronvolt}/\kB$.
Hot gas radiates in X\nobreakdash-rays through inverse Compton scattering and
in the radio through thermal free--free and synchrotron emission. The
nonaxisymmetric minidisk shape means that the electron number density, electron
temperature, and magnetic field strength along the sightline to an observer,
and hence the observed flux and spectrum, may vary periodically as a function
of binary orbital phase in synchronization with the thermal emission.
Determining the exact physical properties of the hot gas and its observational
appearance in X\nobreakdash-rays and the radio requires \ac{RT} that accurately
captures the physics at such high temperatures and the nonaxisymmetry of the
source.

Finally, brightness variation due to minidisk nonaxisymmetry and relativistic
effects is most prominent for edge-on \acp{MBHB}. However, sightlines to such
\acp{MBHB} are also likely to pass through gas and dust around the \acp{MBH}
and in the host galaxy, reducing the observational prospect of the brightness
variation.

\subsection{Observational implications for \ac{LISA} and \ac{PTA} binaries}

After an individual \ac{GW} source is detected by \ac{LISA} or \acp{PTA},
identifying its \ac{EM} counterpart is challenging because the \ac{GW}
localization region can contain a large number of galaxies until just moments
before coalescence. One strategy for improving our chances of finding the host
galaxies of \ac{MBHB} mergers is to search for \acp{MBHB} that would coalesce
in a few years or decades using distinctive \ac{EM} signatures. Being able to
catch \acp{MBHB} early allows us to follow them in \ac{EM} through the last
stages of inspiral, providing us information about the \acp{MBHB} themselves,
their immediate environments, and their host galaxies.

The brightness periodicity our simulation predicts could inform this kind of
preemptive search for \acp{MBHB}. The main observational challenge is that
thermal emission from the minidisks in equal-mass \qty{2e7}{\solarmass}
\acp{MBHB} peaks in the extreme \ac{UV}. Because the color temperatures of the
minidisks should scale with the total mass as $\mathrelp\propto
M_{\su{tot}}^{-1/4}$ and their bolometric luminosity should scale as
$\mathrelp\propto M_{\su{tot}}$, their thermal emission may be easier to detect
in the optical for \qtyrange{\sim e8}{e9}{\solarmass} \acp{MBHB}. Another
obvious challenge is that the search must be performed using telescopes with
high sensitivity, a large field of view, and frequent revisits to the same
field.

\Acp{MBHB} with $M_{\su{tot}}\sim\qty{2e7}{\solarmass}$ coalesce after
\qty{\sim24}{\year} due to \ac{GW} emission. They enter the \ac{LISA} band
during late inspiral when their orbital separations are
$a\lesssim\text{several}\times10\,GM_{\su{tot}}/c^2$ and remain in the band
until merger; therefore, the brightness periodicity we predict is observable
long before the \acp{MBHB} are detected in \acp{GW}. Unfortunately, observing
the Rayleigh--Jeans tail of the thermal emission using X\nobreakdash-ray
instruments is possible only for local \acp{MBHB}; for example, \acp{MBHB} with
a bolometric luminosity of $\mathrelp\sim0.1\,L_{\su E}$, the typical
time-averaged luminosity in \cref{fig:luminosity}, lie above the \acf{ULTRASAT}
flux limit only if they are at $z\lesssim0.2$.

\Acp{MBHB} with $\qty{e8}{\solarmass}\lesssim
M_{\su{tot}}\lesssim\qty{e9}{\solarmass}$ and $a\sim100\,GM_{\su{tot}}/c^2$
should be detectable by \acp{PTA} out to $z\sim1$ as individual sources or as
unresolved sources contributing to the stochastic \ac{GW} background. These
\acp{MBHB} take \qtyrange{\sim e2}{e3}{\year} in the source frame to merge. As
continuous sources, their multimessenger \ac{GW} and \ac{EM} observations can
be carried out independently in time. The bolometric luminosity of their
minidisks is \qtyrange{\sim e45}{e46}{\erg\per\second}; this thermal emission,
if absorbed by large-scale gas and dust surrounding the \acp{MBHB}, can be
reradiated as emission lines and in the \ac{IR}, providing a bright beacon that
points to the presence of \acp{MBH}. The minidisk thermal emission itself is
also bright enough to be seen directly in the optical. Minidisks in \qty{\sim
e9}{\solarmass} \acp{MBHB} is visible out to $z\sim10$ by the \ac{LSST} at the
Vera Rubin Observatory in a \qty{30}{\second} dark-sky exposure in the
$g$\nobreakdash-, $r$\nobreakdash-, and $i$\nobreakdash-bands; conversely,
\qty{\sim e8}{\solarmass} \acp{MBHB} can be detected out to $z\sim4$ only in
the $g$\nobreakdash-band. The observed brightness of the thermal emission
varies with a source-frame period of \qtyrange{\sim0.1}{1}{\year}, which is
longer than the per-field revisit time and shorter than the overall survey
duration. \Ac{ZTF}, a shallower survey by design, may be able to detect the
same \acp{MBHB} in stacked images.

\section{Conclusions}
\label{sec:conclusions}

\Acp{GW} from \acp{MBHB} will likely be detected by the existing \acp{PTA} and
the upcoming \ac{LISA}. \Ac{EM} counterparts are indispensable for constraining
the location and properties of \ac{GW} sources. Predicting the dynamics,
structure, and emission properties of the circumbinary disk and minidisks
hinges on an accurate description of their thermodynamics, but simulations that
treat radiative cooling properly by solving the \ac{RT} equation remain
exceedingly costly.

This article presents the first \acp{RMHD} simulation of the minidisk in an
equal-mass circular \ac{MBHB}. The binary under consideration has total mass
$M_{\su{tot}}=\qty{2e7}{\solarmass}$ and orbital separation
$100\,GM_{\su{tot}}/c^2$; such binaries are moving into the \ac{LISA} band and
will, in a few years or decades, merge in the band. Our simulation results
could be extrapolated to more massive \acp{MBHB} at similar separations; such
binaries are \ac{PTA} sources.

Our setup is centered on one of the \acp{MBH}. We idealize mass supply from the
circumbinary disk as injection of magnetized gas from the boundary, which
gradually builds up a minidisk. In lieu of a moment-based approach, we directly
solve the \ac{RT} equation on a discretized angle grid to more accurately
capture how radiation propagates through an arbitrary gas distribution.

We find that our minidisk is very nonaxisymmetric in its gas and radiation
distributions; consequently, thermal emission from the minidisk is also highly
anisotropic. The two minidisks in the \ac{MBHB} combined act like a lighthouse,
producing a radiation pattern that sweeps around at the binary orbital period.
The total observed brightness of the two minidisks should therefore vary at
half that period. The amplitude of the variation should be a factor of a few
for most observer inclinations. This brightness variation could be especially
useful for identifying \acp{MBH} that are \num{\sim100} gravitational radii
apart, considering that it would be in phase with any detectable \acp{GW}.

The thermal emission from the minidisk, as is typical for disks around
\acp{MBH}, peaks in the extreme \ac{UV}. This observational challenge could be
partially overcome by targeting more massive \acp{MBHB} at higher redshifts
using upcoming all-sky optical transient surveys such as the \ac{LSST}. Besides
thermal emission, hot gas above and below the minidisk could contribute to the
overall luminosity at high energies through inverse Compton scattering, and at
low energies through thermal free--free and synchrotron emission. Detailed
\ac{RT} is essential for ascertaining the physical properties and emission
signatures of this gas.

Our work underlines the importance of proper \ac{RT} treatment to making
\ac{EM} predictions that can be reliably used for observationally locating
\acp{MBHB}. In particular, our prediction of anisotropic minidisk emission
rests squarely on our ability to solve the \ac{RT} equation directly to track
radiation propagation accurately through a nonaxisymmetric minidisk. \Acp{RMHD}
simulations utilizing detailed \ac{RT} are undoubtedly computationally
expensive, but with greater availability of computing power in the future, such
simulations will become the next frontier in theoretically exploring \ac{EM}
counterparts to \acp{GW} from \acp{MBHB}.

\begin{acknowledgments}
The authors thank Jessie Runnoe for the inspiring discussion during her visit.
This material is based upon work supported by the \acf{NASA} under grant
80NSSC19K0319, by the \acf{NSF} under grant AST-1908042, and by the Research
Corporation for Science Advancement under award CS-SEED-2023-008. The Flatiron
Institute is supported by the Simons Foundation. Research cyberinfrastructure
resources and services supporting this work were provided in part by the
\ac{NASA} High-End Computing (\textacsize{HEC}) Program through the \ac{NASA}
Advanced Supercomputing (\textacsize{NAS}) Division at Ames Research Center;
and in part by the Partnership for an Advanced Computing Environment
(\textacsize{PACE}) at the Georgia Institute of Technology, Atlanta, Georgia,
United States of America.
\end{acknowledgments}

\software{Athena++ \citep{2020ApJS..249....4S, 2021ApJS..253...49J}, NumPy
\citep{2020Natur.585..357H}, Matplotlib \citep{2007CSE.....9...90H}}

\begin{appendices}

\section{Quality factors}
\label{sec:quality}

\begin{figure}
\includegraphics{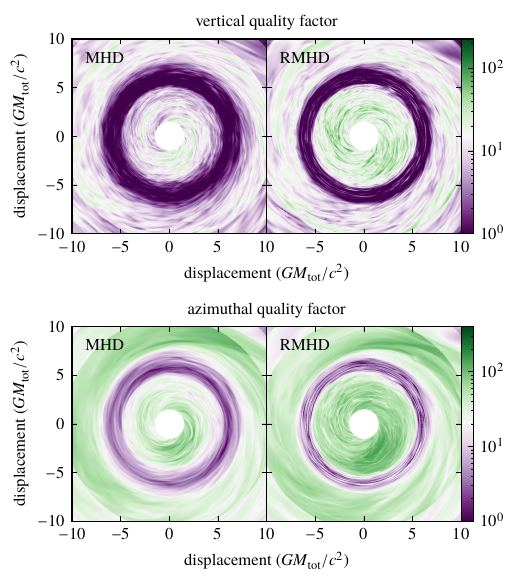}
\caption{Mass-weighted vertical averages of the vertical and azimuthal quality
factors at the end of the \acp{MHD} branch in the left column and the
\acp{RMHD} branch in the right column. The jumps in quality factors near the
corners of each panel are due to changes in mesh refinement level.}
\label{fig:quality}
\end{figure}

We calculate the vertical and azimuthal quality factors at the end of the
\acp{MHD} and \acp{RMHD} branches to determine how well the nonlinear
development of the \ac{MRI} is resolved \citep{2011ApJ...738...84H}:
\begin{align}
Q_\theta &= 2\pi\biggl(\frac{r^3}{{GM_1}}\biggr)^{1/2}
  \frac{\abs{B_\theta}/\rho^{1/2}}{r\Delta\theta}, \\
Q_\phi &= 2\pi\biggl(\frac{r^3}{{GM_1}}\biggr)^{1/2}
  \frac{\abs{B_\phi}/\rho^{1/2}}{r\sin\theta\Delta\phi}.
\end{align}
\Cref{fig:quality} shows their mass-weighted vertical averages:
\begin{align}
\mean{Q_\theta}_{\theta;\rho} &\eqdef
  \int r\,d\theta\,\rho Q_\theta\bigg/\int r\,d\theta\,\rho, \\
\mean{Q_\phi}_{\theta;\rho} &\eqdef
  \int r\,d\theta\,\rho Q_\phi\bigg/\int r\,d\theta\,\rho.
\end{align}
\Acp{MHD} turbulence is deemed properly resolved if $\mean{Q_\theta}\gtrsim15$
and $\mean{Q_\phi}\gtrsim20$ \citep{2013ApJ...772..102H}. The color scales of
the figure are centered on these values for easy visual comparison: green
regions have sufficient resolution.

If we consider the minidisk as a whole, the values of $\mean{Q_\theta}$ and
$\mean{Q_\phi}$ are lower than ideal for resolving \ac{MHD} turbulence. This
could be a consequence of our choice to endow the injected stream with
alternating magnetic field loops, which may promote reconnection in the
minidisk. The dense ring at $r\sim6\,GM_{\su{tot}}/c^2$ has particularly low
quality factors. However, if we consider just the accretion flow at
$r\lesssim5\,GM_{\su{tot}}/c^2$, which produces most of the radiation from the
\acp{RMHD} minidisk, we find $\mean{Q_\theta}\sim3$ and $\mean{Q_\phi}\sim20$
in the \acp{MHD} minidisk, and $\mean{Q_\theta}\sim20$ and
$\mean{Q_\phi}\sim50$ in the \acp{RMHD} minidisk. The \acp{RMHD} minidisk has
higher quality factors than the \acp{MHD} minidisk because its magnetic field
is stronger. Therefore, we expect \acp{MHD} turbulence to be well-developed in
the accretion flow.

\end{appendices}

\ifapj\bibliography{minidisk}\fi
\ifboolexpr{bool{arxiv} or bool{local}}{\bibhang1.25em\printbibliography}{}

\end{document}